\documentclass{ws-procs9x6}

\begin{document}

\title{REALISTIC SHELL-MODEL CALCULATIONS FOR EXOTIC NUCLEI AROUND CLOSED SHELLS\\}

\author{A. COVELLO$^{1,2*}$, L. CORAGGIO$^2$, A. GARGANO$^2$ and N. ITACO$^{1,2}$}  
\address{$^1$ Dipartimento di Scienze Fisiche, Universit\`a di Napoli Federico II, Complesso Universitario di Monte S. Angelo, Via Cintia, I-80126, Napoli, Italy\\
$^2$ Istituto Nazionale di Fisica Nucleare, Sezione di Napoli, Complesso Universitario di Monte S. Angelo, Via Cintia, I-80126 Napoli, Italy\\
$^*$E-mail: covello@na.infn.it\\}

\begin{abstract}
We report on a study of neutron-rich nuclei around doubly magic $^{132}$Sn in terms of the shell model employing a realistic effective interaction derived from the CD-Bonn nucleon-nucleon potential. The short-range repulsion of the bare potential is renormalized by constructing a low-momentum potential,$V_{\rm low-k}$, that is used directly as input for the calculation of the effective interaction. We present results for the four nuclei beyond the $N=82$ shell closure $^{134}$Sn, $^{134}$Sb, 
$^{136}$Sb, and $^{136}$Te. Comparison shows that our results are in very good agreement with the experimental data presently available for these exotic nuclei. We also present our predictions of the hitherto unknown spectrum of $^{136}$Sn.

\end{abstract}


\bodymatter

\section{Introduction}\label{aba:sec1}

The study of neutron-rich nuclei around doubly magic $^{132}$Sn is a subject of special interest, as it offers the opportunity for testing the basic ingredients of shell-model calculations, in particular the two-body effective interaction, when moving toward the neutron drip line. In this context, great attention is currently focused on nuclei beyond the $N=82$ shell closure. The experimental study of these nuclei is very difficult, but in recent years some of them have become accessible to spectroscopic studies. It is therefore challenging to perform shell-model calculations to try to explain the available data as well as to make predictions that may be verified in a not too distant future. 

On these grounds, we have recently studied \cite{Coraggio05,Coraggio06,Covello07a,Covello07b,Simpson07} several nuclei beyond $^{132}$Sn within the framework of the shell model employing realistic effective interactions derived from the CD-Bonn nucleon-nucleon ($NN$) potential \cite{Machleidt01}. 
A main difficulty encountered in this kind of calculations is the strong short-range repulsion contained in the bare $NN$ potential $V_{NN}$, which prevents its direct use in the derivation of the shell-model effective interaction $V_{\rm eff}$.
As is well known, the traditional way  to overcome this difficulty is the Brueckner
$G$-matrix method. Instead, in the calculations mentioned above we have made use of a new approach  \cite{Bogner02} which consists in deriving from $V_{NN}$ a low-momentum potential, $V_{\rm low-k}$, that preserves the deuteron binding energy and scattering phase shifts of $V_{NN}$ up to a certain cutoff momentum $\Lambda$. This is a smooth potential which can be used directly to derive $V_{\rm eff}$, and it has been shown \cite{Bogner02,Covello03} that it provides an advantageous alternative to the use of the $G$ matrix. 

In this paper, we focus attention on $^{134}$Sn, on the two odd-odd Sb isotopes $^{134}$Sb and $^{136}$Sb, and on $^{136}$Te. The first and third nucleus with an $N/Z$ ratio of 1.68 and 1.67, respectively, are at present the most exotic nuclei beyond $^{132}$Sn for which information exists on excited states. We compare our results with the available experimental data and also report our predictions for low-energy states which have not been seen to date. We hope that this may stimulate, and be helpful to, future experiments on these nuclei. With the same motivation, we also present our predictions for the spectrum of the hitherto unknown $^{136}$Sn with four neutrons outside $^{132}$Sn, which makes an N/Z ratio of 1.72.

Following this introduction, in Sec. 2 we give a  brief description of the theoretical framework in which our shell-model calculations are performed while in Sec. 3 we present results and predictions for the nuclei mentioned above. Some concluding remarks are given in Sec. 4. 

\section{Outline of theoretical framework}
The starting point of any realistic shell-model calculation is the free $NN$ potential. There are, however, several high-quality potentials, such as Nijmegen I and Nijmegen II  \cite{Stoks94}, Argonne $V_{18}$ \cite{Wiringa95}, and CD-Bonn \cite{Machleidt01}, which fit equally well ($\chi^2$/datum $\approx 1$) the $NN$ scattering data up to the inelastic threshold. This means that their on-shell properties are essentially identical, namely they are phase-shift equivalent.   
In our shell-model calculations we have derived the effective interaction from the CD-Bonn  potential. This may raise the question of how much our results may depend on this choice of the $NN$ potential. We shall comment on this point later in connection with the $V_{\rm low-k}$ approach to the renormalization of the bare $NN$ potential.

The shell-model effective interaction $V_{\rm eff}$ is defined, as usual, in the following way. In principle, one should solve a nuclear many-body Schr\"odinger equation of the form 
\begin{equation}
H\Psi_i=E_i\Psi_i ,
\end{equation}
with $H=T+V_{NN}$, where $T$ denotes the kinetic energy. This full-space many-body problem is reduced to a smaller model-space problem of the form
\vspace{-.1cm}
\begin{equation}
PH_{\rm eff}P \Psi_i= P(H_{0}+V_{\rm eff})P \Psi_i=E_iP \Psi_i .
\end{equation}
\noindent Here $H_0=T+U$ is the unperturbed Hamiltonian, $U$ being an auxiliary potential introduced to define a convenient single-particle basis, and $P$ denotes the projection operator onto the chosen model space.

As pointed out in the Introduction, we ``smooth out" the strong repulsive core contained in the bare $NN$ potential $V_{NN}$ by constructing a low-momentum  potential $V_{\rm low-k}$. This is achieved by integrating out the \linebreak high-momentum modes of $V_{NN}$ down to a cutoff momentum  $\Lambda$. The integration is carried out with the requirement that the deuteron binding energy and phase shifts of $V_{NN}$ up to $\Lambda$ are preserved by $V_{\rm low-k}$. 
A detailed description of the derivation of $V_{\rm low-k}$ from $V_{NN}$ as well as a discussion of its main features can be found in Refs. \refcite{Bogner02} and \refcite{Covello05}. However, we should mention here that shell-model effective interactions derived from different phase-shift equivalent $NN$ potentials through the $V_{\rm low-k}$ approach lead to very similar results \cite{Covello07b}. In other words, $V_{\rm low-k}$ gives an approximately unique representation of the $NN$ potential.

Once the $V_{\rm low-k}$ is obtained, we use it, plus the Coulomb force for protons, as input interaction  for the calculation of the matrix elements of the shell-model effective interaction. The latter is derived by employing a folded-diagram method, which was previously applied to many nuclei using the $G$ matrix.\cite{Covello01} Since $V_{\rm low-k}$ is already a smooth potential, it is no longer necessary to calculate the $G$ matrix. We therefore perform shell-model calculations following the same procedure as described for instance in  Refs. \refcite{Jiang92} and \refcite{Covello97}, except that the $G$ matrix used there is replaced by $V_{\rm low-k}$. More precisely, we first calculate the so-called $\hat{Q}$-box \cite{Kuo90} including diagrams up to second order in the two-body interaction. The shell-model effective interaction is then obtained by summing up the $\hat{Q}$-box folded diagram series using the Lee-Suzuki iteration method \cite{Suzuki80}. \\

\section{Calculations and results}

In our calculations we assume that $^{132}$Sn is a closed core and let the valence
protons occupy the five levels $0g_{7/2}$, $1d_{5/2}$, $1d_{3/2}$, $2s_{1/2}$, and $0h_{11/2}$ of the 50-82 shell, while for the valence neutrons the model space includes the six levels $0h_{9/2}$, $1f_{7/2}$, $1f_{5/2}$, $2p_{3/2}$,
$2p_{1/2}$, and  $0i_{13/2}$ of the 82-126 shell.

\begin{figure}[h]
\vspace{0.1cm}
\begin{center}
  \begin{tabular}{lr}
     \resizebox{64mm}{!}{\includegraphics{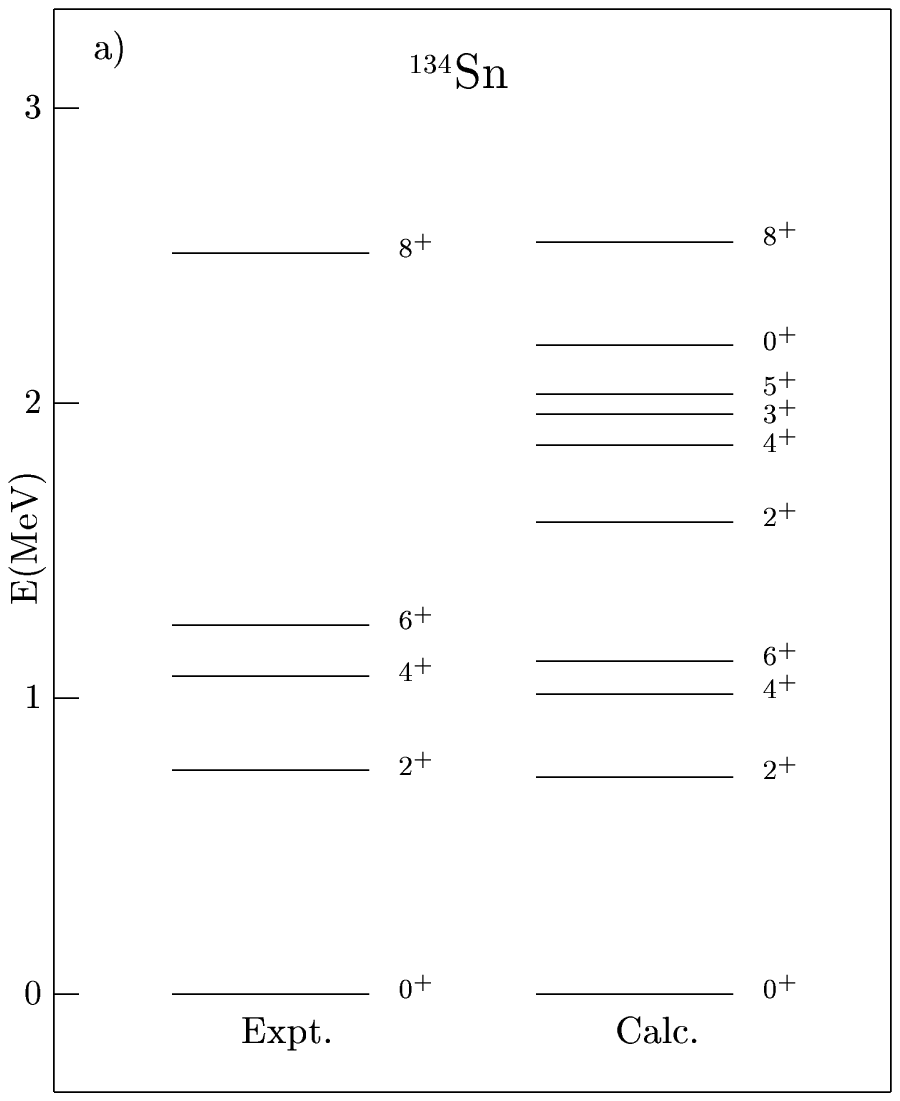}}
\hspace{+0.3cm}
     \resizebox{42mm}{!}{\includegraphics{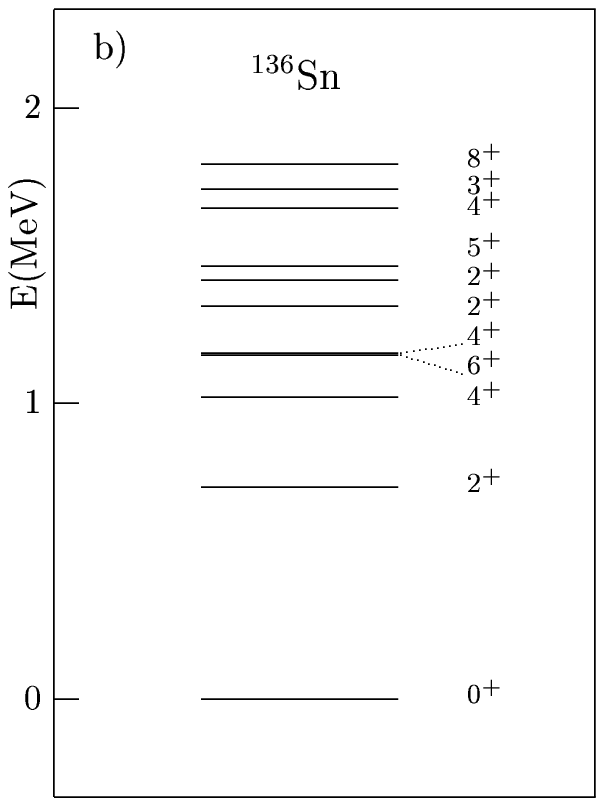}} \\
  \end{tabular}
\end{center}
\caption{a) Experimental and calculated spectrum of $^{134}$Sn. b) Predicted spectrum of $^{136}$Sn.}
\end{figure}

As mentioned in the previous section, the two-body matrix elements of the effective interaction are derived from the CD-Bonn $NN$ potential renormalized through the 
$V_{\rm low-k}$ procedure with a cutoff momentum $\Lambda=2.2$~fm$^{-1}$. The computation of the diagrams included in the $\hat{Q}$-box is performed within the harmonic-oscillator basis using  intermediate states composed of all possible hole states and particle states restricted to the five shells above the Fermi surface. The oscillator parameter used is $\hbar \omega = 7.88$ MeV.
As regards the single-particle energies, they have been taken from experiment. All the adopted values are reported in Ref. \refcite{Coraggio05}. 

In Fig. 1a) we compare the calculated spectrum of $^{134}$Sn with the available experimental data while in Fig. 1b) we show the calculated spectrum of the hitherto unknown $^{136}$Sn up to about 1.8 MeV. From Fig. 1a) we see that while the theory reproduces very well all the observed levels, it also predicts, in between the $6^+$ and $8^+$ states, the existence of five states with spin $\leq 5$.  Clearly, the latter could not be seen from the $\gamma$-decay of the $8^+$ state populated in the spontaneous fission experiment of Ref. \refcite{Korgul00}.

Recently, the $B(E2;0^+ \rightarrow 2_1^+)$ value in $^{134}$Sn has been measured \cite{Beene04} using Coulomb excitation of neutron-rich radioactive ion beams. We have calculated this $B(E2)$ with an effective neutron charge of $0.70\,e$. We obtain $B(E2;0^+ \rightarrow 2_1^+)$ = 0.033 $e^2$b$^2$, in excellent agreement with the experimental value 0.029(4) $e^2$b$^2$. 

A comparison between figures 1a) and 1b) shows that the three lowest calculated states, $2^+$, $4^+$, and $6^+$, lie at practically the same energy in $^{134}$Sn and
$^{136}$Sn. In the latter nucleus, however, above the $6^+$ level there are seven states in an energy interval of about 650 keV. This pattern is quite different from that predicted for the spectrum of $^{134}$Sn, where a rather pronounced gap (about 0.5 MeV wide) exists between the $6^+$ state and the next excited state with $J^\pi=2^+$.

Let us now come to $^{134}$Sb. In Fig. 2 we show the energies of the first eight calculated states, which are the members of the $\pi g_{7/2} \nu f_{7/2}$  multiplet, and compare them with the eight lowest-lying experimental states.  
The wave functions of these states are characterized  by 
very little configuration mixing, the percentage of the leading component having a minimum value of 88\% for the $J^{\pi}=2^{-}$ state 
while ranging from 94\% to 100\% for all other states.

\begin{figure}[h]
\begin{center}
\psfig{file=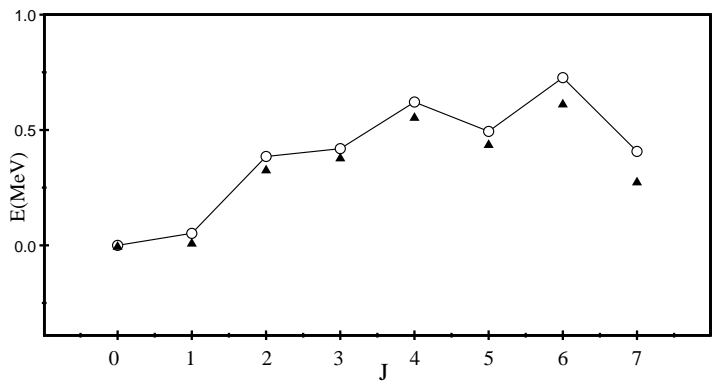,width=3.7in}
\end{center}
\caption{Proton-neutron $\pi g_{7/2} \nu f_{7/2}$  multiplet in $^{134}$Sb. The theoretical results are represented by open circles and the experimental data by solid triangles.} 
\end{figure}

We see that the agreement between theory and experiment is very good, the  discrepancies being in the order of a few tens of keV for most of the states. The largest discrepancy occurs for the $7^{-}$ state, which lies at about 130 keV 
above its experimental counterpart. It is an important outcome of our calculation  that we predict  almost the right spacing between the $0^{-}$ ground state and first excited $1^{-}$ state. In fact, the latter has been observed at 13 keV excitation energy, our value being 53  keV. 

An analysis of the various terms which contribute to our effective neutron-proton interaction has evidenced the crucial role of core-polarization effects, in particular those arising from $1p-1h$ excitations, in reproducing the correct behavior of the multiplet. A detailed discussion of this 
point can be found in Ref. \refcite{Coraggio06}. 

As regards $^{136}$Sb, with two more valence neutrons, its ground state was identified as $1^-$ in the early  $\beta$-decay study of Ref. \refcite{Hoff97} while the spectroscopic study of Ref. \refcite{Mineva01} led to the observation of a $\mu$s isomeric state, which was tentatively assigned a spin and parity of $6^-$. Very recently \cite{Simpson07}, new experimental information has been obtained on the $\mu$s isomeric cascade, leading to the identification of two more excited states. 
 
\begin{figure}[h]
\begin{center}
\psfig{file=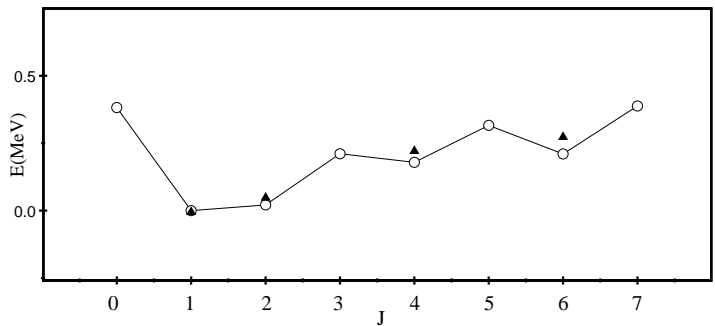,width=3.7in}
\end{center}
\caption{Low-lying levels in $^{136}$Sb. The theoretical results are represented by open circles and the experimental data by solid triangles.} 
\end{figure}

This achievement is at the origin of our realistic shell-model calculation for this nucleus \cite{Simpson07}, whose results we are now going to present. In Fig.~3, we show the four observed levels together with the calculated yrast states having angular momentum from $0^-$ to $7^-$, which all arise from the $\pi g_{7/2} \nu (f_{7/2})^{3}$ configuration. These states may be viewed as the evolution of the $\pi g_{7/2} \nu f_{7/2}$  multiplet in $^{134}$Sb. From Fig. 3 we see that the agreement between theory and experiment is very good, the largest discrepancy being about 70 keV.
 
An important piece of information is provided by the measured half life of the $6^-$ state, from which a $B(E2;6^-\rightarrow 4^-)$ value of 170(40) $e^2$ fm$^4$ is extracted. Using effective proton and neutron charges of 1.55 $e$ and 0.70~$e$, respectively, we obtain the value 131 $e^2$ fm$^4$, which compares very well with experiment. It is worth mentioning that these values of the effective charges have been consistently used in our previous calculations for nuclei in the $^{132}$Sn region \cite{Covello07b}. 

Finally, we turn to the two-proton, two-neutron nucleus $^{136}$Te, which in recent years has been the subject of great experimental and theoretical interest \cite{Radford02,Terasaki02,Shimizu04,Brown05}. Particular attention 
has been focused on the first $2^+$ state, which shows a significant drop in energy as compared to $^{134}$Te, and on  the  $B(E2; 0^{+} \rightarrow 2^{+})$  value, recently measured using Coulomb excitation of radioactive ion beams \cite{Radford02}. 

\begin{figure}[h]
\begin{center}
\psfig{file=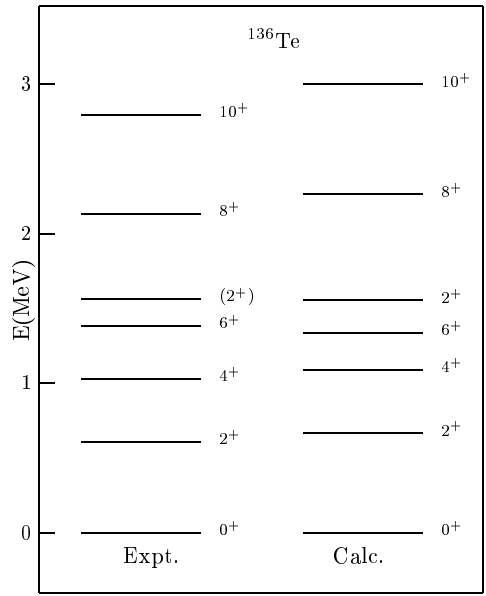,width=2.5in}
\end{center}
\caption{Experimental and calculated spectrum of $^{136}$Te.} 
\end{figure}

From Fig. 4 we see that also in this case the calculated spectrum reproduces very well the experimental one.

As regards the $B(E2; 0^{+} \rightarrow 2^{+})$, the experimental value reported in Ref. \refcite{Radford02} is 0.103(15) $e^2$ b$^2$, to be compared with our value 0.180 $e^2$ b$^2$. The latter has been obtained with the same effective charges used for $^{136}$Sb. We see that the calculated value overestimates the experimental one by a factor of 1.7. However, two new measurements have been recently performed which yield values about 20\% \cite{Mach07} and 50\% \cite{Baktash} higher than that of Ref. \refcite{Radford02}.
Both values are closer to our prediction.

\section{Concluding remarks}

We have presented here the results of a shell-model study of neutron-rich nuclei, focusing attention on the four nuclei $^{134}$Sn, $^{134}$Sb, $^{136}$Sb, and 
$^{136}$Te, with an $N/Z$ ratio of 1.68, 1.63, 1.67, and 1.61, respectively. We have also presented our predictions for the spectrum of the hitherto unknown  $^{136}$Sn with four neutrons outside $^{132}$Sn, which makes an N/Z ratio of 1.72.  In this connection, it may be recalled that the $N/Z$ ratio of the last stable Sn isotope, $^{124}$Sn, is 1.48. 

In our study we have made use of a realistic two-body effective interaction derived  from the CD-Bonn $NN$ potential, the short-range repulsion of the latter being renormalized by use of the low-momentum potential $V_{\rm low-k}$. It is worth emphasizing that no adjustable parameter appears in our calculation of the effective interaction. 

We have shown that all the experimental data presently available for these nuclei are well reproduced by our calculations. These data, however, are still rather scanty
and we hope that our study may stimulate further experimental efforts to gain more spectroscopic information on exotic nuclei around  $^{132}$Sn.

\section*{Acknowledgments}

This work was supported in part by the Italian Ministero dell'Istruzione, 
dell'Universit\`a e della Ricerca (MIUR). \\


\noindent
\begin{thebibliography}{50}

\bibitem{Coraggio05} L. Coraggio, A. Covello, A. Gargano, and N. Itaco, 
{\it Phys. Rev. C} {\bf72}, 057302 (2005).
\bibitem{Coraggio06} L. Coraggio, A. Covello, A. Gargano, and N. Itaco, 
{\it Phys. Rev. C} {\bf73}, 031302(R) (2006).
\bibitem{Covello07a} A. Covello, L. Coraggio, A. Gargano, and N. Itaco, {\it Eur. Phys. J. ST } {\bf 150}, 93 (2007).
\bibitem{Covello07b} A. Covello, L. Coraggio, A. Gargano, and N. Itaco, {\it Prog. Part. Nucl. Phys.}  {\bf 59}, 401 (2007).
\bibitem{Simpson07} G. S. Simpson, J. C. Angelique, J. Genevey, J. A. Pinston, A. Covello, A. Gargano, U. K\"oster, R. Orlandi, and  A. Scherillo, 
{\it Phys. Rev. C} \  {\bf 76},  041303(R) (2007).
\bibitem{Machleidt01} R. Machleidt, {\it Phys. Rev. C} {\bf 63}, 024001 (2001).
\bibitem{Bogner02} S. Bogner, T. T. S. Kuo, L. Coraggio, A. Covello, and N. Itaco,  
{\it Phys. Rev. C} {\bf 65}, 051301(R) (2002).
\bibitem{Covello03} A. Covello, in {\em  Proc. Int. School of Physics ``E. Fermi", Course CLIII}, eds. A. Molinari, L. Riccati, W. M. Alberico, and M. Morando (IOS Press, Amsterdam, 2003), p. 79.
\bibitem{Stoks94} V. G. J. Stoks, R. A. M. Klomp, C. P. F. Terheggen, and J. J. de Swart,  
{\it Phys. Rev. C} {\bf 49}, 2950 (1994).
\bibitem{Wiringa95} R. B.Wiringa, V. G. J. Stoks, and R. Schiavilla, {\it Phys. Rev. C} {\bf 51}, 38 (1995).
\bibitem{Covello05} A. Covello, L. Coraggio, A. Gargano, and N. Itaco, {\it Journal of Physics: Conference Series} {\bf 20}, 137 (2005).
\bibitem{Covello01} A. Covello, L. Coraggio, A. Gargano, and N. Itaco, {\it Acta Phys. Pol. B} {\bf 32}, 871 (2001), and references therein.
\bibitem{Jiang92} M. F. Jiang, R. Machleidt, D. B. Stout, and T. T. S. Kuo,  {\it Phys. Rev. C} {\bf 46}, 910 (1992). 
\bibitem{Covello97} A. Covello, F. Andreozzi, L. Coraggio, A. Gargano , T. T. S. Kuo, and A. Porrino, {\it Prog. Part. Nucl. Phys.} {\bf 38}, 165 (1997).
\bibitem{Kuo90} T. T. S. Kuo and E. Osnes, \textit{Lecture Notes in Physics}, Vol. 364, (Springer-Verlag, Berlin, 1990).
\bibitem{Suzuki80} K. Suzuki and S. Y. Lee,  {\it Prog. Theor. Phys.} {\bf 64}, 2091 (1980).
\bibitem{Korgul00} A. Korgul {\it et al.}, {\it Eur. Phys. J. A} {\bf 7}, 167 (2000).
\bibitem{Beene04} J. R. Beene {\it et al.}, {\it Nucl. Phys. A} {\bf 746}, 471c (2004).
\bibitem{Hoff97} P. Hoff, J. P. Omtved, B Fogelberg, H. Mach, and M. Hellstr{\"o}m,
{\it Phys. Rev. C} {\bf 56}, 2865 (1997).
\bibitem{Mineva01} M. N. Mineva {\it et al.}, {\it Eur. Phys. J. A} {\bf 11}, 9 (2001).
\bibitem{Radford02} D. C. Radford {\it et al.}, {\it Phys. Rev. Lett.} {\bf88}, 222501 (2002).
\bibitem{Terasaki02} J. Terasaki, J. Engel, W. Nazarewicz, and M. Stoitsov,
{\it Phys. Rev. C} {\bf 66}, 054313 (2002).
\bibitem{Shimizu04} N. Shimizu, T. Otsuka, T. Mizusaki, and M. Honma, 
{\it Phys. Rev. C} {\bf 70}, 054313 (2004). 
\bibitem{Brown05}B. A. Brown, N. J. Stone, R. J. Stone, I. S. Towner, and M. Hjorth-Jensen, {\it Phys. Rev. C} {\bf 71}, 044317 (2005).
\bibitem{Andreozzi97} F. Andreozzi, L. Coraggio, A. Covello, A. Gargano, T. T. S. Kuo, and  A. Porrino, {\it Phys. Rev. C} {\bf 56}, R16 (1997).
\bibitem{Coraggio02} L. Coraggio, A. Covello, A. Gargano, and N. Itaco, 
{\it Phys. Rev. C}  {\bf 65}, 051306(R) (2002).
\bibitem{Mach07} H. Mach {\it et al.}, contribution to these Proceedings.
\bibitem{Baktash} C. Baktash (private communication).


\end{thebibliography}
\end{document}